\documentclass[a4paper]{jpconf}
\usepackage{graphicx}

\newcommand{\pT}{p_\mathrm{T}}
\newcommand{\kt}{k_\mathrm{T}}
\newcommand{\et}{E_\mathrm{T}}
\newcommand{\gev}{\mathrm{GeV}}
\newcommand{\gevc}{\mathrm{GeV}/c}

\begin{document}
\title{Jets in 200 GeV p+p and d+Au collisions from the STAR experiment at RHIC}

\author{Jan Kapit\'an (for the STAR Collaboration)}

\address{Nuclear Physics Institute ASCR, Na Truhlarce 39/64, 18086 Praha 8, Czech Republic}

\ead{kapitan@rcf.rhic.bnl.gov}

\begin{abstract}
Full jet reconstruction in heavy-ion collisions is a promising tool for the quantitative study of properties of the dense medium produced at RHIC. Measurements of d+Au collisions are important to disentangle initial state nuclear effects from medium-induced $\kt$ broadening and jet quenching. Study of jet production and properties in d+Au in combination with similar studies in p+p is an important baseline measurement needed to better understand heavy-ion results.
We present mid-rapidity inclusive jet $\pT$ spectra and di-jet correlations ($\kt$) in 200~GeV p+p and d+Au collisions from the 2007-2008 RHIC run. We discuss the methods used to correct the data for detector effects and for background in d+Au collisions.

\end{abstract}

\section{Introduction}
\label{intro}
Jets are remnants of hard-scattered partons, which are the fundamental objects of pQCD. At RHIC, they can be used as a probe of the hot and dense matter created in heavy ion collisions~\cite{MP,EB}. To quantify the signals observed in heavy ion collisions in comparison to p+p collisions, it is necessary to measure the cold nuclear matter effects in systems such as d+Au. There are several ways to describe cold nuclear matter effects. Partonic rescattering~\cite{vitev} leads to $\kt$ broadening and to modification of jet $\pT$ spectra. Nuclear parton distribution functions~\cite{eps} influence jet $\pT$ spectra, but the effects at mid-rapidity are not expected to be very strong.

\section{Jet reconstruction}
\label{jets}
This analysis is based on $\sqrt{s_\mathrm{NN}} = 200~\gev$ p+p and d+Au data from the STAR experiment, recorded during RHIC run 8 (2007-2008). The Barrel Electromagnetic Calorimeter (BEMC) detector is used to measure the neutral component of jets, and the Time Projection Chamber (TPC) detector is used to measure the charged component of jets. In the case of a TPC track pointing to a BEMC tower, its momentum is subtracted from the tower energy to avoid double counting (electrons, MIP and possible hadronic showers in the BEMC). To reduce possible BEMC backgrounds, the jet neutral energy fraction is required to be within $(0.1,0.9)$. An upper $\pT < 15~\gevc$ cut was applied to TPC tracks due to uncertainties in TPC tracking performance at high-$\pT$ in run 8 (under further investigation). 

The Beam Beam Counter detector, located in the Au nucleus fragmentation region, was used to select the 20\% highest multiplicity events in d+Au collisions. The acceptance of TPC and BEMC together with experimental details (calibration, primary vertex position cuts) limit the jet fiducial acceptance to $|\eta|<0.55 \; (R=0.4), |\eta|<0.4 \; (R=0.5)$, where $R$ is the resolution parameter used in jet finding.

Recombination jet algorithms kt and anti-kt from the FastJet package~\cite{fj} are used for jet reconstruction. To subtract the background, a method based on active jet areas~\cite{bgsub} is applied event-wise: $\pT^{Rec} = \pT^{Candidate} - \rho \cdot A$, with $\rho$ estimating the background density per event and $A$ being the jet active area. Due to the asymmetry of the colliding d+Au system, the background is asymmetric in $\eta$. This dependence was fit with a linear function in $\eta$ and included in the background subtraction procedure. 

To study detector effects, Pythia 6.410 and GEANT detector simulations were used. Jet reconstruction was run at MC hadron level (PyMC) and at detector level (PyGe). To study effects of the d+Au background, a sample with added background (PyBg) was created by mixing PyGe events with 0-20\% highest multiplicity d+Au events (minimum bias online trigger).

\section{Nuclear $\kt$ broadening}

Azimuthal correlations of jets in di-jet events in p+p and d+Au can provide information on nuclear $\kt$ broadening. To increase di-jet yield, BEMC high tower (HT) online trigger was required (one tower with $\et > 4.3~\gev$), for both p+p and d+Au. A cut $\pT > 0.5~\gevc$ applied for tracks and towers to reduce background and the jet finding was run with $R=0.5$. 

\begin{figure}[htb]
\begin{minipage}[h]{0.31\textwidth}    
\includegraphics[width=\textwidth]{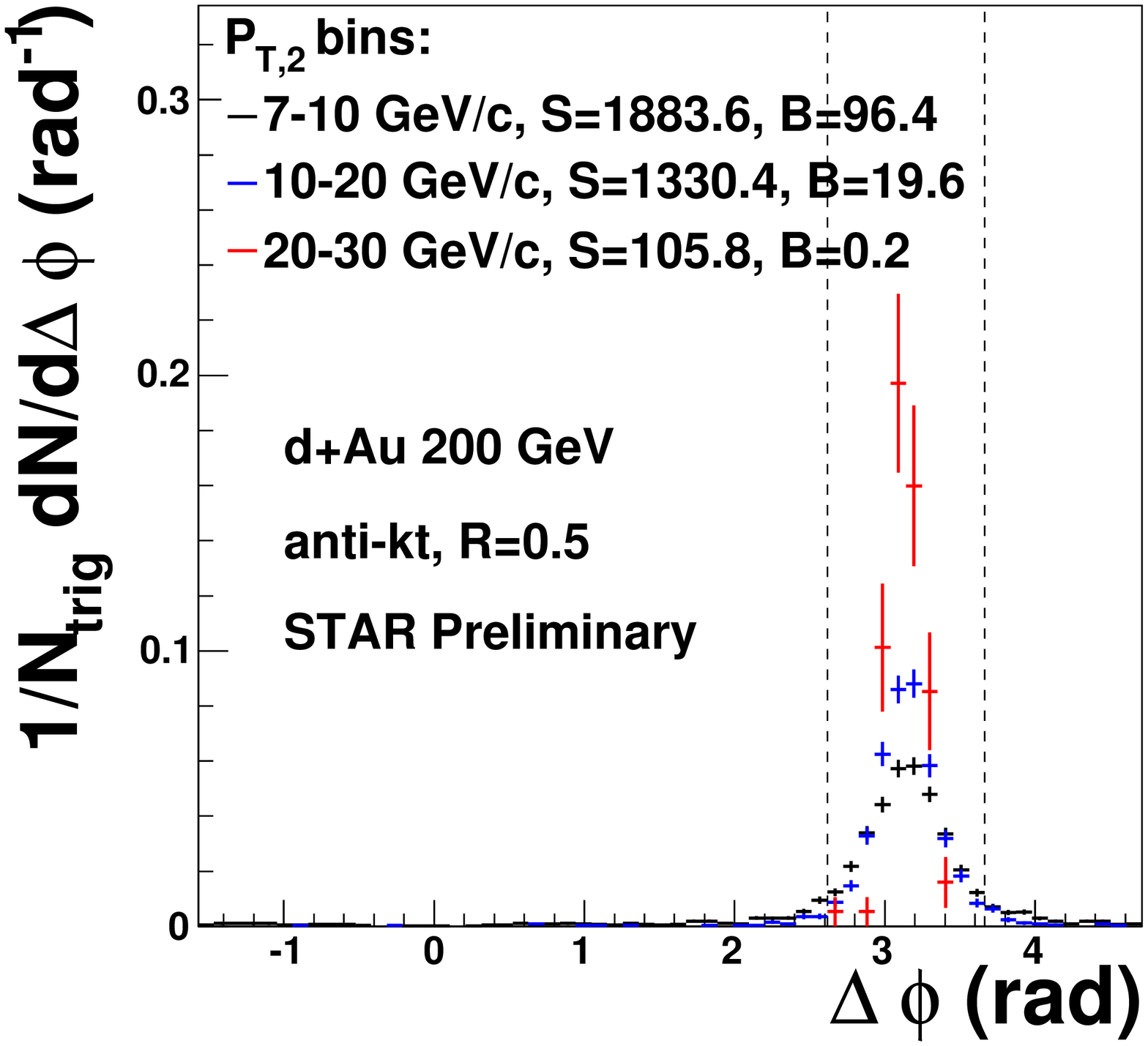}
\vspace{-1.0cm}
\caption{\label{fig:dphi}Di-jet azimuthal correlations in 3 $p_\mathrm{T,2}$ bins.}
\end{minipage}
\hfill
\begin{minipage}[h]{0.65\textwidth}  
\includegraphics[width=\textwidth]{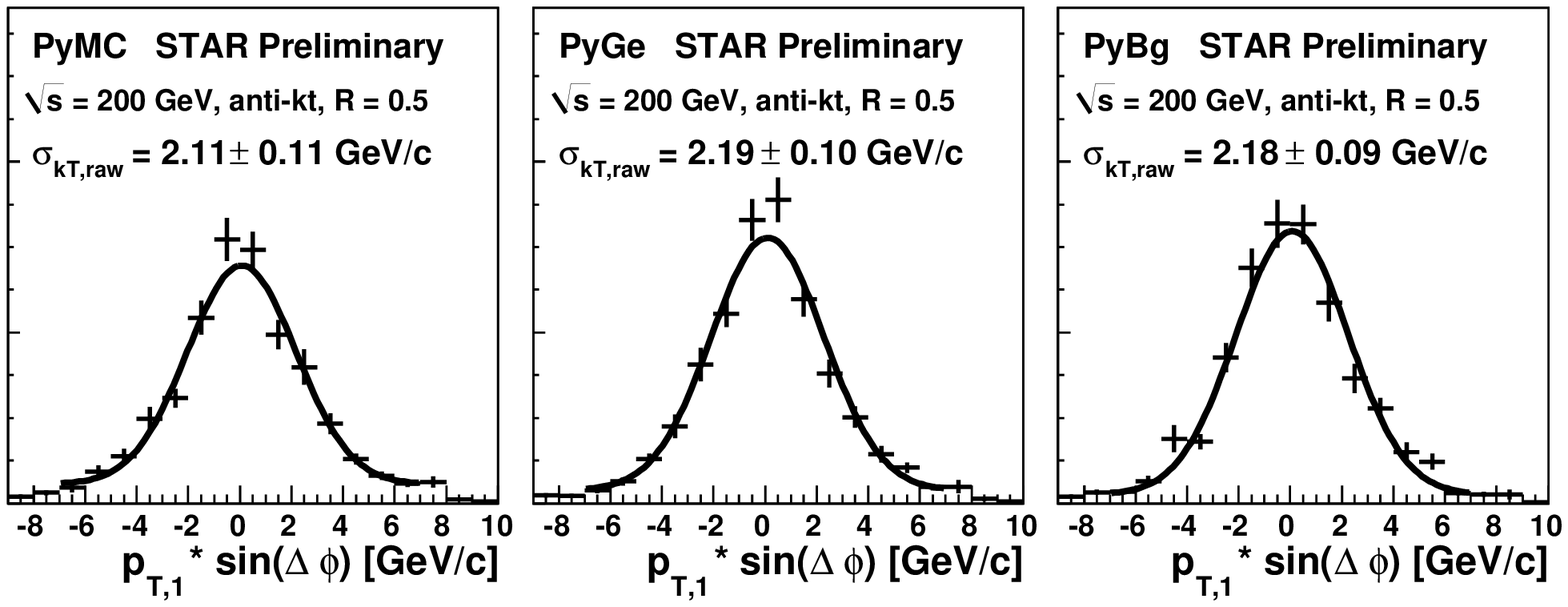}
\vspace{-0.55cm}
\caption{\label{fig:ktsimu}Distributions of $k_\mathrm{T,raw} = p_\mathrm{T,1} \sin(\Delta\phi)$ for simulation ($10 < p_\mathrm{T,2} < 20~\gevc$).}
\end{minipage}
\end{figure}

To select a clean di-jet sample two highest energy jets ($p_\mathrm{T,1} > p_\mathrm{T,2}$) in each event were used, with $p_\mathrm{T,2} > 10~\gevc$. Observed di-jet signal in d+Au collisions is shown in Figure~\ref{fig:dphi}. Distributions of $k_\mathrm{T,raw} = p_\mathrm{T,1} \sin(\Delta\phi)$ were constructed for di-jets and Gaussian widths, $\sigma_{k_\mathrm{T,raw}}$, were obtained for the two jet algorithms and two ($10 - 20~\gevc$, $20 - 30~\gevc$) $p_\mathrm{T,2}$ bins. 

Figure~\ref{fig:ktsimu} shows that the widths are similar for PyMC, PyGe and PyBg distributions, therefore detector and background effects are negligible, due to the interplay between jet $\pT$ and $\Delta\phi$ resolutions. Figure~\ref{fig:ktdata} shows an example of the $k_\mathrm{T,raw}$ distributions for data. 
The Gaussian fit to p+p data is not ideal and the precise shape of the distribution is under study. RMS widths of these distributions have therefore been checked and they agree with the sigma widths of the fits. The values extracted from the Gaussian fits are $\sigma_{k_\mathrm{T,raw}}^{p+p} = 2.8 \pm 0.1~\mathrm{(stat)}~\gevc$ and $\sigma_{k_\mathrm{T,raw}}^{d+Au} = 3.0 \pm 0.1~\mathrm{(stat)}~\gevc$. Possible nuclear $\kt$ broadening therefore seems rather small.

The systematic errors on $\sigma_{k_\mathrm{T,raw}}$ coming from a weak dependence on the $|\Delta\phi - \pi|$ cut for back-to-back di-jet selection (varied between 0.5 and 1.0) and differences between the various selections ($p_\mathrm{T,2}$ range, jet algorithm) are estimated to be $0.2~\gevc$. Additional uncertainty comes from Jet Energy Scale (JES), which will be discussed in more detail in section~\ref{systematics}.

\section{Inclusive jet spectra}
\label{spectra}
10M 0-20\% highest multiplicity d+Au events with a minimum bias online trigger were used for this study. Anti-kt algorithm was used with $R = 0.4$, $\pT > 0.2~\gevc$ cut was applied to tracks and towers. 
The jet $\pT$ spectrum is normalized per event and the high multiplicity of d+Au events also guarantees the trigger efficiency is independent of the $\pT$ of the hard scattering. Therefore, no correction related to trigger is needed for jet $\pT$ spectra.

\begin{figure}[htb]
\begin{minipage}[h]{0.58\textwidth}
\includegraphics[width=\textwidth]{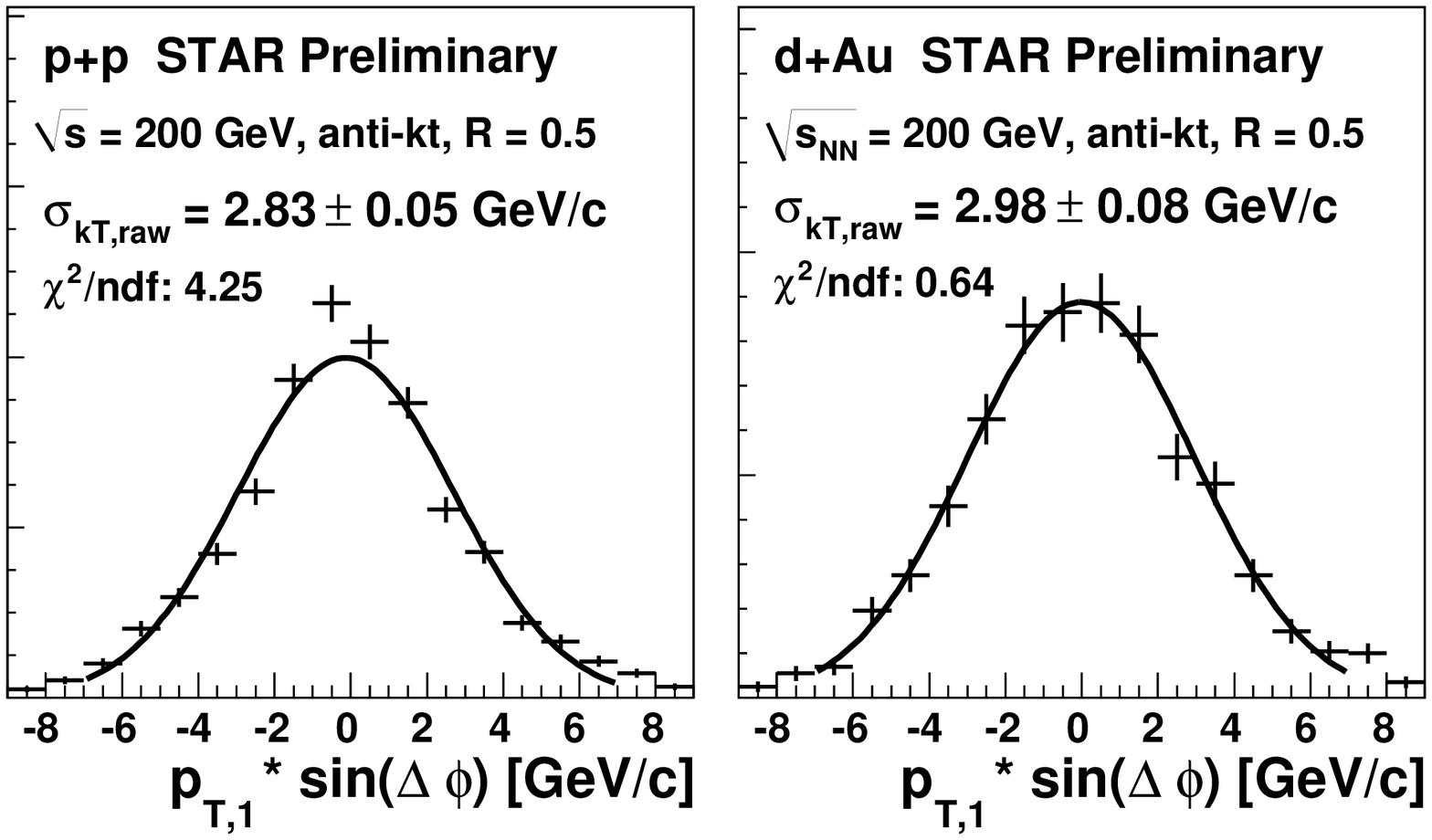}
\vspace{-0.75cm}
\caption{\label{fig:ktdata}Distributions of $k_\mathrm{T,raw}$ for p+p, d+Au ($10 < p_\mathrm{T,2} < 20~\gevc$).}
\end{minipage}
\hfill
\begin{minipage}[h]{0.38\textwidth}    
\includegraphics[width=\textwidth]{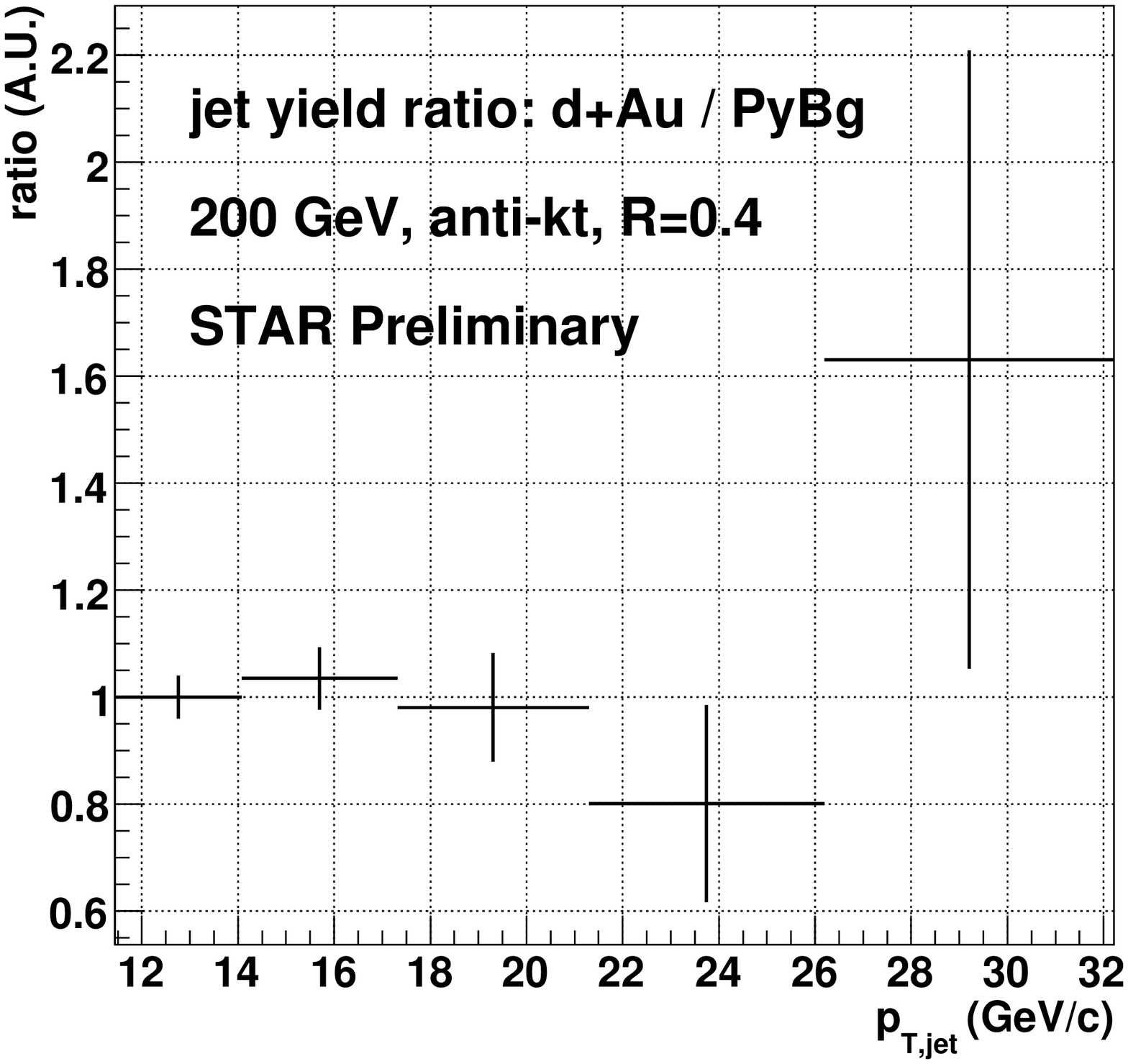}
\vspace{-1.1cm}
\caption{\label{fig:ratio}Ratio of jet $\pT$ spectra between d+Au data and simulation.}
\end{minipage}
\end{figure}

A bin-by-bin correction is used to correct for detector effects (tracking and tower efficiency, unobserved neutral energy, jet $\pT$ resolution). It is based on the generalized efficiency, constructed as the ratio of PyMC to PyBg jet $\pT$ spectra. These correction factors are then applied to measured jet $\pT$ spectrum. As the detector effects on jet $\pT$ spectrum differ substantially depending on jet $\pT$ spectrum shape, the shapes have to be consistent between PyBg and measured jet $\pT$ spectra. Figure~\ref{fig:ratio} shows that this is indeed the case.

As the jet $\pT$ spectrum is very sensitive to the jet energy scale, an additional correction was applied here to account for the lower TPC tracking efficiency in d+Au compared to that from the used p+p Pythia simulation. The d+Au efficiency was determined by simulating single pions and embedding them at the raw detector level into real d+Au minimum bias events. 
The tracking efficiency in the Pythia simulation was then artificially lowered, prior to jet finding at PyGe and PyBg level, so that it matches the one obtained from d+Au embedding. 

To compare per event jet yield in d+Au to jet cross section in p+p collisions, an input from MC Glauber study is needed. In the following, we have used $\langle N_\mathrm{bin} \rangle = 14.6 \pm 1.7$ for 0-20\% highest multiplicity d+Au collisions and $\sigma_\mathrm{inel,pp} = 42~\mathrm{mb}$. These factors were used to scale p+p jet cross section measured previously by the STAR collaboration~\cite{ppjetprl} using Mid Point Cone (MPC) jet algorithm with $R = 0.4$.
The resulting d+Au jet $\pT$ spectrum is shown in Figure~\ref{fig:spectrum} together with scaled p+p jet spectrum. Within systematic uncertainties, the d+Au jet spectrum shows no significant deviation from the scaled p+p spectrum.
The discrepancy between simulation and embedding was only discovered after the d+Au spectrum was presented in the conference, and it was corrected only for these proceedings. The difference is however smaller than the combined systematic and statistical uncertainties, the conclusions therefore remain the same.

\section {Systematic uncertainties}
\label{systematics}
\begin{figure}[htb]
\includegraphics[width=0.9\textwidth]{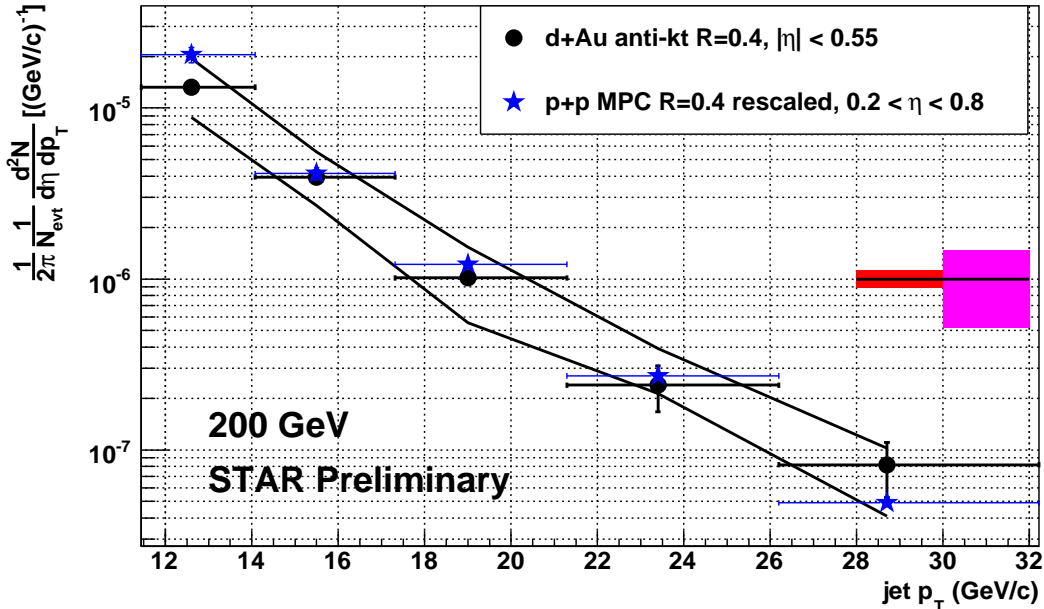}
\vspace{-0.5cm}
\caption{\label{fig:spectrum}Jet $\pT$ spectrum from d+Au collisions compared to scaled p+p spectrum~\cite{ppjetprl}. Red box indicates uncertainty of $\langle N_\mathrm{bin} \rangle$, black lines indicate JES uncertainty in d+Au and the magenta box shows the total systematic uncertainty of p+p measurement (including JES uncertainty).}
\end{figure}

The JES uncertainty dominates the uncertainties of d+Au measurement and is marked by the black lines in Figure~\ref{fig:spectrum}. It has two components: 5\% on the neutral jet component (BEMC calibration uncertainty) and 10\% on the charged jet component (uncertainty of TPC tracking efficiency in jets in d+Au collisions). Embedding of jets into real d+Au events at raw detector level will allow to decrease this uncertainty in the future.
As JES uncertainty is expected to be largely correlated between run 8 p+p and d+Au data, we plan to measure jet $\pT$ spectrum in run 8 p+p collisions to decrease these uncertainties and be able to construct jet $R_\mathrm{dAu}$. 

Caution is needed due to the use of different jet algorithms and different pseudorapidity acceptances in Figure~\ref{fig:spectrum}. Although neither of these effects can change our conclusion, the same algorithm and the same acceptance have to be used in p+p and d+Au to construct jet $R_\mathrm{dAu}$.

\section{Summary}
\label{summary}

Di-jet $\kt$ widths were measured in 200 GeV p+p and d+Au collisions: $\sigma_{k_\mathrm{T,raw}}^{p+p} = 2.8 \pm 0.1~\mathrm{(stat)}~\gevc$, $\sigma_{k_\mathrm{T,raw}}^{d+Au} = 3.0 \pm 0.1~\mathrm{(stat)}~\gevc$. No significant broadening due to Cold Nuclear Matter effects was observed.
Jet $\pT$ spectrum from minimum bias 200 GeV d+Au collisions is consistent with scaled p+p jet spectrum within systematic uncertainties. Precise tracking efficiency determination from jet embedding and cross section measurement in run 8 p+p data will allow to construct jet $R_\mathrm{dAu}$. 

\section*{Acknowledgement}
\label{acknowledgement}

This work was supported in part by grants LC07048 and LA09013 of the Ministry of Education of the Czech Republic and by the grant SVV-2010-261 309.

\end{document}